\documentclass[aps,prl,reprint,superscriptaddress]{revtex4-1}
\bibstyle{apsrev}	

\usepackage{graphicx}
\usepackage{color}
\usepackage{dcolumn}

\begin{document}

\title{The Sub-band Structure of Atomically Sharp Dopant Profiles in Silicon}

\author{Federico~Mazzola}
\affiliation{Center for Quantum Spintronics, Department of Physics, Norwegian University of Science and Technology, NO-7491 Trondheim, Norway }
\affiliation{SUPA, School of Physics and Astronomy, University of St Andrews, St Andrews, Fife KY16 9SS, UK.}
\author{Chin-Yi Chen}
\affiliation{Department of Electrical and Computer Engineering, Purdue University, West Lafayette, IN 47907, USA}
\author{Rajib Rahman}
\affiliation{Department of Electrical and Computer Engineering, Purdue University, West Lafayette, IN 47907, USA}
\affiliation{School of Physics, The University of New South Wales, Sydney, New South Wales 2052, Australia}
\author{Xie-Gang Zhu}
\affiliation{Science and Technology on Surface Physics and Chemistry Laboratory, Mianyang 621907, China}
\author{Craig~M.~Polley}
\affiliation{MAX IV Laboratory, PO Box 118, S-22100 Lund, Sweden}
\author{Thiagarajan Balasubramanian}
\affiliation{MAX IV Laboratory, PO Box 118, S-22100 Lund, Sweden}
\author{Phil~D.~C.~King}
\affiliation{SUPA, School of Physics and Astronomy, University of St Andrews, St Andrews, Fife KY16 9SS, UK.}
\author{Philip Hofmann}
\affiliation{Department of Physics and Astronomy, Interdisciplinary Nanoscience Center (iNANO), University of Aarhus, 8000 Aarhus C, Denmark}
\author{Jill~A.~Miwa}
\affiliation{Department of Physics and Astronomy, Interdisciplinary Nanoscience Center (iNANO), University of Aarhus, 8000 Aarhus C, Denmark}
\author{Justin~W.~Wells}
\email[]{quantum.wells@gmail.com}
\affiliation{Center for Quantum Spintronics, Department of Physics, Norwegian University of Science and Technology, NO-7491 Trondheim, Norway }

\date{\today}

\maketitle

\textbf{The downscaling of silicon-based structures and proto-devices has now reached the single atom scale, representing an important milestone for the development of a silicon-based quantum computer \cite{Fuechsle:2012,Weber:2012,Zwanenburg:2013,Watson:2018}. One especially notable platform for atomic scale device fabrication is the so-called Si:P $\delta$-layer, consisting of an ultra dense and sharp layer of dopants within a semiconductor host. Whilst several alternatives exist, phosphorus dopants in silicon have drawn the most interest, and it is on this platform that many quantum proto-devices have been successfully demonstrated \cite{Fuechsle:2010,Tettamanzi:2017,Broome:2018,Koch:2019}.   Motivated by this, both calculations and experiments have been dedicated to understanding the electronic structure of the Si:P $\delta$-layer platform \cite{Carter:2009,Lee:2011b,Carter:2011,Drumm:2013a,Miwa:2013,Miwa:2014a,Mazzola:2014a,Mazzola:2014b}.  In this work, we use high resolution angle-resolved photoemission spectroscopy (ARPES)  to reveal the structure of the electronic states which exist because of the high dopant density of the Si:P $\delta$-layer. In contrast to published theoretical work, we resolve three distinct bands, the most occupied of which shows a large anisotropy and significant deviation from simple parabolic behaviour. We investigate the possible origins of this fine structure, and conclude that it is primarily a consequence of the dielectric constant being large (ca.\ double that of bulk Si) \cite{Ristic:2004}. Incorporating this factor into tight binding calculations leads to a major revision of band structure; specifically, the existence of a third band, the separation of the bands, and the departure from purely parabolic behaviour. This new understanding of the bandstructure has important implications for quantum proto-devices which are built on the Si:P $\delta$-layer platform.}

\begin{figure*}
\centering
\includegraphics [width=5in]{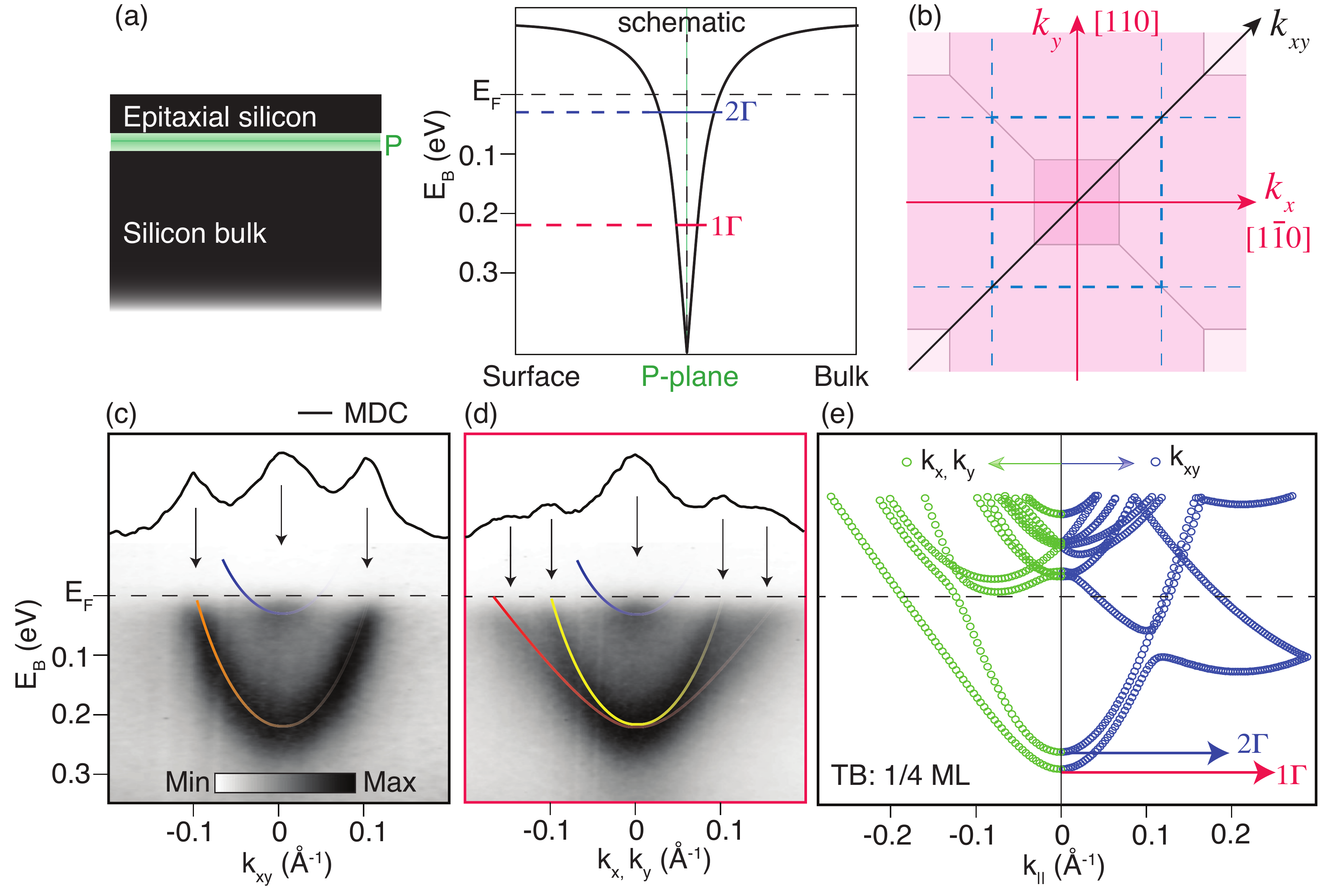}
\caption{\textbf{Si:P $\delta$-layers, ARPES measurements and the sub-band structure.} (a) Schematic of the $\delta$-layer sample and the symmetric doping potential centred around the position of the P dopant plane. The strong potential created by the P atoms confines the conduction band, developing a 2D electronic structure which has been understood to be comprised of two states, labeled $1\Gamma$ (red) and $2\Gamma$ (blue). (b) The Brillouin Zone (BZ) of silicon showing the directions relevant for this work. The blue dashed lines show the 2D BZ, and the pink shading and lines shows a 2D slice through the 3D BZ of silicon at the relevant value of $k_z$ for this work (darkest: 1st bulk BZ). The high symmetry directions $k_x$, $k_y$ (red axes) and $k_{xy}$ (black axis) are also indicated. (c) ARPES acquisition along the $k_{xy}$ direction showing only two states. A momentum distribution curve (MDC) at the Fermi level ($E_F$) shows no presence of additional states. (d) ARPES measurement along the $k_{x}$ (or $k_{y}$) direction showing the existence of three states, (marked with blue, red and yellow lines to serve as a guide for the eye). (e) Tight-binding calculations (following Ref.\ \onlinecite{Lee:2011b}) showing the $1\Gamma$ and $2\Gamma$ states dispersing across $E_F$ (with $E_F$ adjusted such that the minimum of the dispersion matches with the data in panels (c) and (d)) along the $k_{xy}$ and $k_x$ (or $k_y$) directions (blue and green markers respectively). As discussed in the main text, the agreement between the TB calculation and the ARPES data in panels (c) and (d) is unsatisfactory.}\label{Fig1}
\end{figure*}

Si:P $\delta$-doping offers potential for the realization of true atomic-scale components for quantum computer applications, whilst retaining compatibility with the simple processing, stability and technological relevance of silicon.  Understanding, manipulating and controlling the properties of Si:P $\delta$-layers, has therefore been the centre of an intense research effort, however, a real understanding of the electronic structure has remained elusive. Density functional theory (DFT) calculations and ARPES recently shed new light on these systems, giving the first glimpse of their electronic structure \cite{Carter:2009, Drumm:2013a, Carter:2013, Miwa:2013}: the metallic nature of Si:P $\delta$-layers was believed to originate from two nearly-parabolic states, called $1\Gamma$ and $2\Gamma$, dispersing across the Fermi level ($E_{F}$) as a consequence of the strong electronic confinement created by the P dopants in the semiconducting Si bulk (see Fig.\ \ref{Fig1}(a)). The energy separation of these states, which is called valley-splitting \cite{Miwa:2014a}, together with their many-body interactions \cite{Mazzola:2014a} is responsible for transport properties in this material  system and ultimately the function  of Si:P $\delta$-layer based quantum electronic devices.

In this work, we show that important details of the electronic bandstructure were previously reported incorrectly. We reveal the presence of additional anisotropic electronic states crossing $E_F$, resolved only for specific directions in the BZ (see Fig.\ \ref{Fig1}(b)-(d)). Whilst in the diagonal direction ($k_{xy}$) only two electronic states can be seen, along the axial directions ($k_x$ and $k_y$), a clear $3$-band structure is resolved which has not been predicted.  The original $1\Gamma$ appears to actually consist of two sub-bands, indicated by the red and yellow parabolae in Fig.\ \ref{Fig1}(d). The presence of three states across $E_F$ cannot be reconciled with published DFT \cite{Carter:2009,Carter:2011,Drumm:2013a}. and tight-binding (TB) calculations \cite{Lee:2011b}. This discrepancy is also seen in our TB calculations (Fig.\ \ref{Fig1}(e) plus details in the Methods section \cite{suppl}), where only two bands, instead of three, are responsible for the metallic properties of the system. This opens an interesting question about the origin of the sub-band structure resolved by ARPES, as all the states which contribute to this sub-band structure are expected to contribute to the transport properties of Si:P $\delta$-layers.

We examine some ingredients which have previously been ignored, such as spin orbit coupling (SOC), the role of the dielectric constant, $\epsilon$, and an asymmetric doping profile, to explain the origin of this sub-band structure. We show that $\epsilon$ is dramatically increased in the vicinity of the high density dopant layer, and that this causes additional states originally predicted to be well above $E_F$ (as in Fig.\ \ref{Fig1}(e)) to become occupied. 

\begin{figure*}
\centering
\includegraphics [width=4.5in]{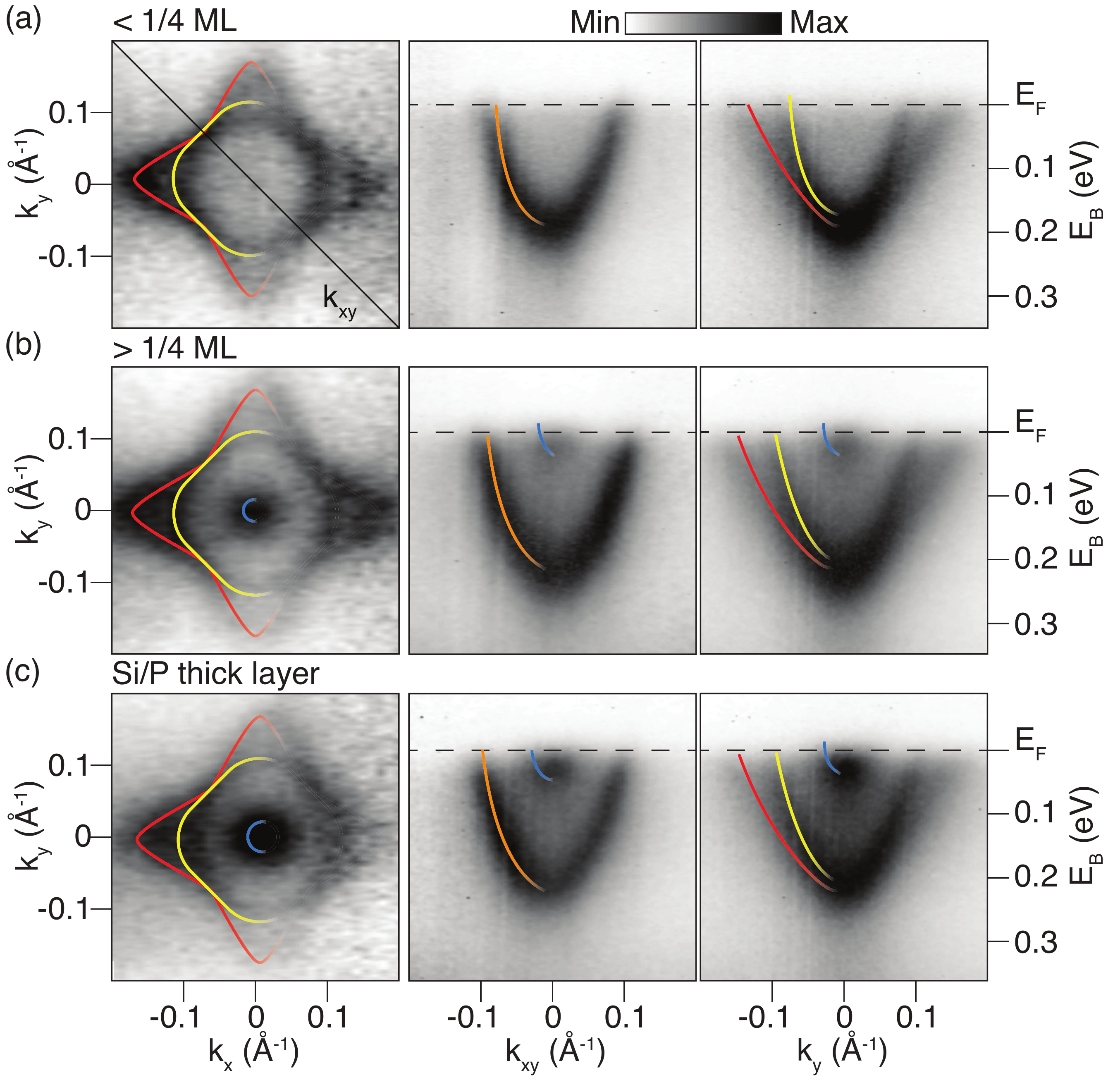}
\caption{\textbf{Ubiquity of the sub-band structure in differently grown doping profiles.} Fermi surface and electronic structures along the $k_{xy}$ and $k_{y}$ directions for a Si:P $\delta$-layer grown with a P concentration (a) slightly less than $1/4$~monolayer (ML), (b) more than $1/4$~ML and (c) by depositing P and Si together such to form a thicker layer ($\approx 1.5$~nm) with a similar ($\approx25$$\%$) doping concentration. Blue, red and yellow lines to serve as a guide for the eye. }\label{Fig2}
\end{figure*}

Before further discussing the origin of the fine structure of Si:P $\delta$-layers, we first present a qualitative discussion of the electronic structure and the parameters to which it is sensitive. First of all, the available calculations have predicted two nearly parabolic bands with a valley splitting of $\approx$30~meV \cite{Carter:2009,Lee:2011b,Carter:2011,Drumm:2013a}. It is worth noting that this value is somewhat controversial, and depends on a number of parameters within the model, such as the order/disorder of the dopants \cite{Carter:2011}. Our measurements reveal the presence of \emph{three} bands, of which the most occupied bands have a valley splitting which is too small to resolve experimentally (i.e.\ $<35$~meV). In other words, the observed valley splitting is either small, or zero. Furthermore, in the axial $k_x$ and $k_y$ directions, the dispersion of the most occupied band deviates significantly from parabolic behaviour, whereas in the diagonal direction (labelled $k_{xy}$) the dispersion of these bands is very close to parabolic, and they appear to either be degenerate, or to have a very small separation. The state with its minimum closest to the Fermi level (i.e.\ the blue parabola in Fig.\ \ref{Fig1}(c,d)) is separated from the other bands by $\approx$220~meV, which is very large compared to our TB calculated valley splitting. In other words, it is unclear which of the three bands (if any) correspond to the calculated $1\Gamma$ and $2\Gamma$. It is especially unclear whether the additional band  is split off from $1\Gamma$, or whether it actually corresponds to the calculated $2\Gamma$ (thereby implying that the least occupied band actually has another origin). In any case, it is clear that the calculated bandstructure deviates significantly from the experimentally observed bandstructure.  

\begin{figure*}
\centering
\includegraphics [width=7in]{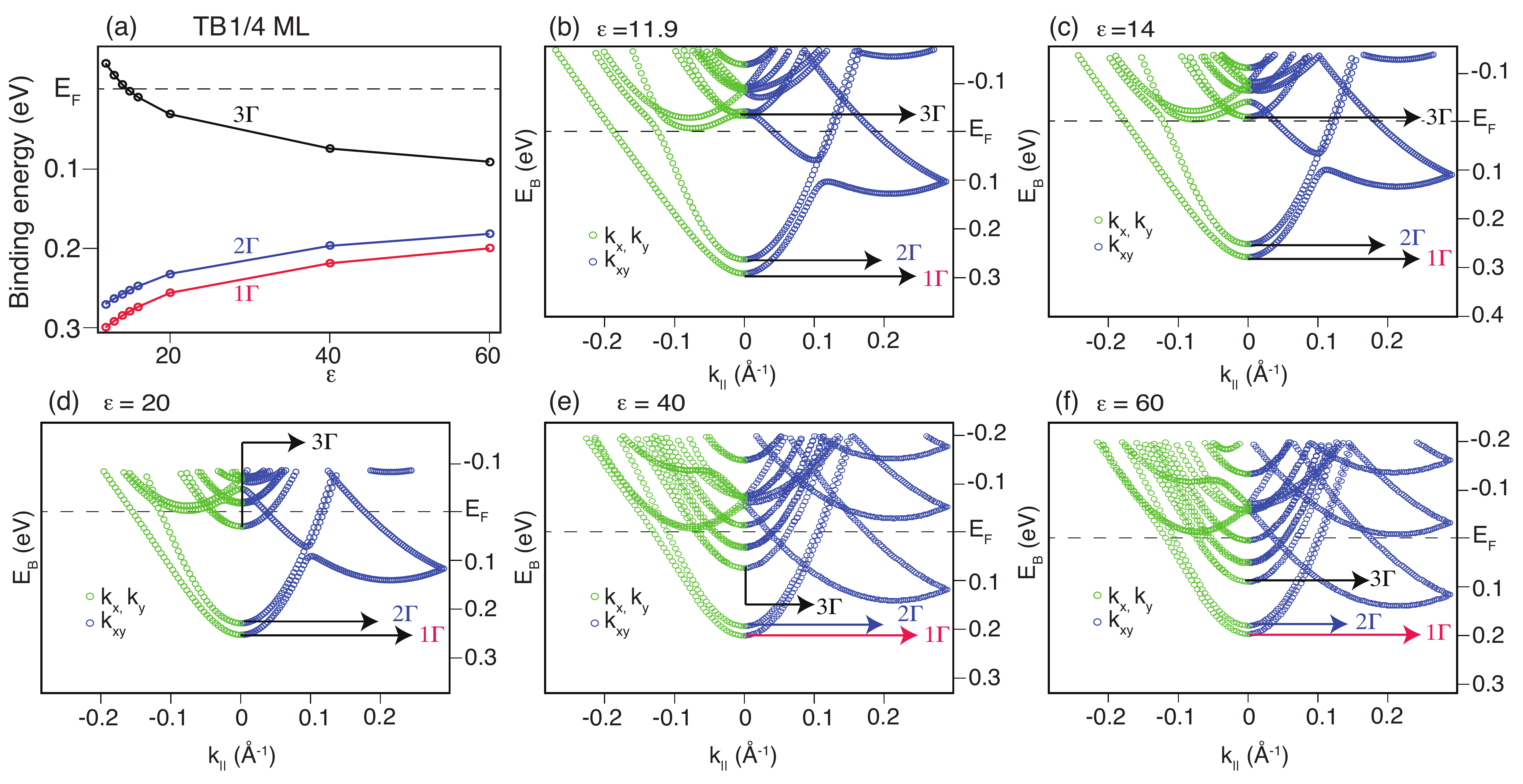}
\caption{\textbf{Role of the relative dielectric constant in the calculated bandstructure.} (a) Energy levels of $1\Gamma$ (red), $2\Gamma$ (blue) and $3\Gamma$ (black) as a function of dielectric constant. By increasing $\epsilon$, keeping all other factors constant, the energy $1\Gamma$-$2\Gamma$ remains approximately constant. However, the separation $1\Gamma$-$3\Gamma$ significantly shrinks and the $3\Gamma$ state starts to cross $E_F$.  $\epsilon\approx20$ gives the best agreement with experimental data. (b)-(f) The $k$ resolved TB calculations are shown for the axial $k_x$ and $k_y$ direction (green $\circ$) and for the diagonal $k_{xy}$ direction (blue $\circ$) for; (b) $\epsilon=11.9$, (i.e.\ for weakly doped Si, and consistent with previous calculations), (c) $\epsilon=14$, (d) $\epsilon=20$, (d) $\epsilon=40$ and (d) $\epsilon=60$.}\label{Fig3new}
\end{figure*}

In principle, symmetry breaking in one form or another could give rise to additional bands. More specifically, the in-plane dopant order/disorder is not well known in practice, but \textit{is} thought to influence the electronic structure \cite{Carter:2013}. In any case, whilst the electronic structure is clearly influenced by the symmetry and ordering of the dopants, this is not able to account for the observed bandstructure \cite{suppl}.  

This notion that dopant ordering is not a significant factor is supported empirically: We have prepared three different $\delta$-layer samples, in which the dopant ordering is dissimilar, however the electronic structure remains very similar. The spectra in Fig.\ \ref{Fig2}(a) correspond to a `standard' single-dose Si:P $\delta$-layer with $<$$1/4$ of a monolayer (ML) of P dopants in an almost atomically sharp plane. Fig.\ \ref{Fig2}(b) corresponds to a similarly sharp `double dose'  with dopant density $>$$1/4$~ML and increased disorder \cite{McKibbin:2014}.  Fig.\ \ref{Fig2}(c) corresponds to a `thick' (1.5~nm) region with a similar (i.e.\ $\approx$25$\%$) doping concentration (see Supplementary Material for further details \cite{suppl}).

The measured band structure is very similar for all three preparations: The electronic structures of Fig.\ \ref{Fig2} map onto each other very well, needing only a small shift of the bands, to account for the different degrees of doping. Perhaps the most significant difference is that the least occupied band is shifted above the Fermi level when the doping density is lowest. The similarity of the bandstructure for these three different growth methods indicates that dopant ordering cannot be responsible for the mysterious 3rd band.

Symmetry breaking (specifically, the breaking of in-plane inversion symmetry by the dopants in the $\delta$-layer) together with spin-orbit coupling (SOC) will lift degeneracy and thus give rise to bands which are non-degenerate with respect to their spin \cite{winklerbook,Rotenberg:1999}. This could lead to $1\Gamma$ having two branches, with no splitting at $k_{\parallel}=0$, anisotropic splitting at larger $k_{\parallel}$, and a bandstructure which qualitatively matches the ARPES measurements. However, the expected energy splitting due to SOC is about $120\times$ smaller than the observed energy separation of the two most occupied bands \cite{suppl,Ferdous:2018,Ferdous:2018a}.  We therefore discount SOC as a possible origin of the observed bandstructure.

The modification of $\epsilon$ in the vicinity of the $\delta$-layer is surprisingly important and can significantly influence the electronic structure. For the moderate doping densities found in semiconducting Si wafers, $\epsilon$ is typically considered to be independent of dopant concentration, but for the extreme doping around the $\delta$-layer, this view is no longer valid \cite{Ristic:2004}.  

For degenerately doped semiconductors, as the dopant density is increased, a subsequent increase in the susceptibility of the material, and thus in its dielectric constant, can be expected \cite{Bethin:1974,Masumi:1978,Dhar:1985}. Following the method of Risti\'{c} \textit{et al.}\ \cite{Ristic:2004}, it is possible to estimate $\epsilon$ as a function of dopant density ($N_D$, in units of cm$^{-3}$), for phosphorous dopants in Si:
\begin{equation} \label{Eq_2}
\epsilon(N_D)=\epsilon_{intrinsic}+\frac{1.6\times10^{âˆ-19}N_D}{1+1.2\times10^{âˆ-21}N_D}
\end{equation}

For the case of Si:P $\delta$-layers, the 2D doping concentration is known to be close to $1/4$~ML.  Whilst there is some uncertainty involved in converting the 2D concentration to a 3D concentration, this problem has been addressed previously  (see, for example Refs.\ \onlinecite{Polley:2013a,McKibbin:2014,Suzuki:2007}).  Previous works suggest that in the vicinity of the $\delta$-layer $N_D$ is $2\times10^{20}$~cm$^{-3}$ \cite{McKibbin:2014,Suzuki:2007}, and our own previous work \cite{Polley:2013a} suggests that whilst the peak concentration is $\approx 6\times10^{20}$~cm$^{-3}$, we agree that the average concentration within 2-4~nm of the layer is $\approx2\times10^{20}$~cm$^{-3}$. Using Eqn.\ \ref{Eq_2}, we can therefore estimate that for $N_D=2\times10^{20}$~cm$^{-3}$, $\epsilon\approx38$, however at the peak of $N_D\approx6\times10^{20}$~cm$^{-3}$ it is conceivable that near to the dopant plane, $\epsilon$ could be as high as 70.  In any case, it is clear that the dielectric constant maybe be several times higher than the value for weakly doped bulk Si.

The dependence of the band structure on $\epsilon$ can be understood in terms of screening: If we consider the $\delta$-layer as a metal sheet sandwiched within a semiconducting host, then the out-of-plane electric field in the vicinity of the dopant plane will depend on $\epsilon$ because a larger dielectric constant is associated with more efficient screening. This means that the quantum well is less confined than previously thought (for example, Ref.\ \onlinecite{Lee:2011b}) and as a result, the splitting between some of the bands is reduced. This qualitative understanding is readily confirmed by TB calculations (see Fig.\ \ref{Fig3new}).  Interestingly, by increasing $\epsilon$ the $1\Gamma-2\Gamma$ valley-splitting stays roughly constant. However, an additional parabolic band minimum (identified from the calculations as $3\Gamma$) is pulled down towards the Fermi level. For $\epsilon >16$,   $3\Gamma$ starts to become occupied, and TB calculation looks more similar to the measured bandstructure. As $\epsilon$ continues to increase, additional parabolic band minima may also be pulled below the Fermi level, and for $\epsilon>40$, a 4th state also becomes partially occupied.

The energy of the $1\Gamma$, $2\Gamma$ and $3\Gamma$ minima are plotted in Fig.\ \ref{Fig3new}(a). By comparison with our experimental data, we find best agreement for $\epsilon \approx 20$: i.e.\ depending on the preparation, we generally observe the minimum of the most occupied state ($1\Gamma$) to be at $E_B\approx220$~meV, and the minimum of the least occupied state (now assigned as $3\Gamma$) to be at $E_B\approx40$~meV. We also find good agreement with the TB electronic dispersions of Fig.\ \ref{Fig3new}(b)-(d); The TB calculation indicates that the two most occupied bands (now assigned as $1\Gamma$ and $2\Gamma$) are parabolic, with a small energy separation (30~meV) at $k_\parallel$=0 and in the diagonal $k_{xy}$ direction, but have an increasing non-parabolic behaviour with increasing $k$ in the axial $x$ and $y$ directions. 

Based on ARPES measurements and TB calculations, we conclude that the enhancement of $\epsilon$ due to the high dopant density is the origin of the additional electronic structure, however, some discrepancies with previous work remain. In previous studies on similar samples  \cite{Miwa:2014a}, the  1$\Gamma$-2$\Gamma$ valley-splitting was reported to be $\approx 130$~meV, whereas in this work, we conclude that the 1$\Gamma$-2$\Gamma$ splitting is $<$30~meV. Due to the lower sample quality and data quality in the earlier work, this was unresolvable and mistaken for a single band, and therefore the previously reported splitting of 130~meV presumably corresponded instead to the 1$\Gamma$ (or 2$\Gamma$) to 3$\Gamma$ splitting. On the other hand, this splitting is still small compared to the TB calculations and ARPES measurements here in which the 2$\Gamma$-3$\Gamma$ splitting is $\approx$200~meV. Whilst it is not possible to give a definitive explanation for this, we propose that it is most likely also a consequence of poorer sample quality; it is known that 1/4~ML dopant activation is only achievable when the Si surface is pristine, and that imperfections act to reduce this number. We therefore consider that the previously reported 130~meV valley splitting presumably corresponds to the (1$\Gamma$ or 2$\Gamma$) to 3$\Gamma$ splitting of a sample with a lower doping concentration that used in the current work.

Finally, we reiterate that a revision of the electronic structure of Si:P is necessary in which three nearly parabolic bands, $1\Gamma$, $2\Gamma$ and $3\Gamma$ all cross $E_F$, contrary to only two bands as previously thought. Jointly, all three of these bands must be responsible for the transport properties of the system \cite{Polley:2012a,Polley:2013a} and carrier density. Importantly, the valley-splitting, i.e.\ separation between $1\Gamma$ and $2\Gamma$ seems to be relatively robust against variations in the sample preparation and can be estimated to be $\le35$~meV. This is an important result which also shows how the properties of a device built upon a Si:P $\delta$-layer platform are not dramatically affected by changes in the growth, but instead are reliable due to the robustness of the valley-splitting. Indeed, since the valley-splitting in devices built from the Si:P $\delta$-layer platform affects the lifetime of carriers  \cite{Hsueh:2014},  its correct  value and interpretation is important for quantum device performance. Similarly, the presence of a third band crossing $E_F$ will have significant consequences for Si:P $\delta$-layer based quantum devices, and hence it is important that this is taken into consideration when developing device structures.

\textbf{Acknowledgements:} This work was partly supported by the Research Council of Norway through its Centres of Excellence funding scheme, project number 262633, `QuSpin', and through the Fripro program, project number 250985 `FunTopoMat' the by the VILLUM FONDEN through the Centre of Excellence for Dirac Materials (Grant No. 11744).  J.~A.~M.~acknowledges funding support from the Danish Council for Independent Research, Natural Sciences under the Sapere Aude program (Grant No. DFF-6108-00409) and the Aarhus University Research Foundation. P.~D.~C.~K. acknowledges financial support from The Royal Society.

%\bibliography{library6}


\end{document}